\documentstyle[aps, prb, epsf, multicol]{revtex}
\begin{document}

\title{Dephasing time of disordered two-dimensional electron gas in 
modulated magnetic fields} 
\author{Xiao-Bing Wang}
\address{The Abdus Salam International Centre for Theoretical Physics,
P.O. Box 586, 34100 Trieste, Italy\\}
%\date{\today}
\maketitle

\begin{abstract}
The dephasing time of disordered two-dimensional electron gas in modulated
magnetic field ${\mathbf{H}}=(0,0,H/\cosh^2((x-x_0)/\delta))$ is studied. In the
weak inhomogeneity limit where $\delta$ is much larger than the linear size of the
sample, $\tau_\phi^{-1}$ is proportional to $H$. In the
strong inhomogeneity limit, it is shown that the dependence 
is quadratic, $\tau_\phi^{-1}=D\left( \frac e {\hbar c} \right) ^2 H^2\delta^2$. 
In the intermediate regime, a crossover between these two limits
occurs at $H_c=\frac{\hbar c}{4e}\delta^{-2}$. It is demonstrated that
the origin of the dependence of $\tau_\phi$ on $H$ lies in the nature of corresponding
single particle motion. A semiclassical Monte Carlo algorithm is developed to study the dephasing
time, which is of qualitative nature but efficient in uncovering the dependence of
$\tau_\phi$ on $H$ for arbitrarily complicated magnetic field modulation.
Computer simulations support analytical results. The crossover from linear to  
quadratic dependence is then generalized to situation with magnetic field modulated
periodically in one direction with zero mean, and it is argued that this crossover 
can be expected for a large class of modulated magnetic fields.


\end{abstract}

\begin{multicols}{2}
\narrowtext

\section{Introduction}
Dephasing is one of the key elements in the study of quantum coherent phenomena in
mesoscopic systems. Coupling with environment suppresses the quantum
interference of electrons. The phase breaking time, or dephasing time is 
the characteristic time beyond which the phase coherence is lost. Although
a static magnetic field does not destroy all quantum effects, it may introduce a cutoff
to the interference effect. Dephasing of disordered electron gas by coupling with a
uniform magnetic field has been 
studied\cite{alt,hik,kaw} in the early stage of weak-localization theory. In the
absence of spin-orbit scattering, the magnetic field suppresses the
weak-localization effect and leads to a positive magnetoconductance, which has
been observed(for a review, see [4]). The dephasing rate due to coupling with 
uniform field turns out to
be proportional to the field amplitude. These results can also be established by
qualitative considerations according to Khmelnitskii\cite{khm}.

In recent years there has been increasing interest\cite{pee} in hybrid semiconductor
systems both for fundamental understanding and for the potentiality of
enhancing the functionality of the devices. 
The disposition of superconducting\cite{geim} or magnetic\cite{ye} microstructures 
on the surface of heterostructure with a two-dimensional
electron gas(2DEG) may produce inhomogeneous magnetic field which influences 
electron motion locally.  Some interesting consequences of the modulated magnetic
fields have already been reported. It was shown\cite{kato} that the
effect of electron-electron Umklapp scattering
can be observed in a 2DEG at a GaAs/Ga$_{1-x}$Al$_{x}$As interface by imposing a
spatially alternating magnetic field normal to the 2DEG plane.
Kubrak et al\cite{kub} fabricated a few different types of hybrid
ferromagnetic-semiconductor devices, which allow them to study how these different
modulated magnetic fields influence the transport properties of 2DEG.  
And magnetoresistance oscillations due to internal Landau band
structure of a 2D electron system in periodic magnetic field are
observed\cite{edm}. Theoretically, Peng\cite{peng}
calculated the transport properties in a parabolic channel exposed to a
periodically modulated magnetic field. Gumbs and Zhang\cite{gz} developed a
magneto-transport theory
for the magnetoconductivity of a square lattice in a periodically modulated magnetic
field, and predicted some anomalies due to commensurability effects.  Recently
Matulis and Peeters\cite{mp} studied the semiclassical magnetoresistance in weakly
modulated magnetic
fields. They considered the case when the field is periodic in one dimension and
with zero mean. In the limit of small magnetic field amplitude, it is shown that the
contribution of the magnetic modulation to magnetoresistance increases as $B^{3/2}$
in the diffusive limit, while increases linearly in $B$ in the ballistic limit.

The magnetoresistance in the diffusive limit is well described by standard
weak-localization theory. It arises from the suppression of magnetic field on the
cooperon propagator which represents the interference of time-reversal
trajectories. Taking the magnetic field into consideration, the cooperon
propagator satisfies\cite{alt}, 

\begin{equation}
\left [-i\omega+D \left (-i\nabla - \frac {2e}{c}{\mathbf{A}}({\mathbf{r}}) \right
)^2\right ] C({\mathbf{r}},{\mathbf{r}}';\omega)=\frac 1 \tau
\delta({\mathbf{r}}-{\mathbf{r}}'), 
\label{diff}
\end{equation}
which may be viewed as an imaginary-time Schr$\ddot{\rm o}$dinger equation with
parameters properly replaced. For the uniform field situation, the solution of the
corresponding Schr$\ddot{\rm o}$dinger equation is the well-known Landau
levels. The dephasing time was found to be\cite{alt,hik,kaw} 

\begin{equation}
\frac 1 {\tau^u_\phi}=\frac {4DeH}{\hbar c},
\label{dep1}
\end{equation}
where the superscript $u$ represents result for uniform field.

In this paper we first study the dephasing time of disordered 2DEG due to coupling with
the following modulated magnetic field 

\begin{equation}
{\mathbf{H}}=(0,0,H/\cosh^2((x-x_0)/\delta)). 
\label{field}
\end{equation}
The corresponding Schr$\ddot{\rm o}$dinger equation has been solved exactly by
Hud$\acute{\rm a}$k\cite{hud}, where it was shown that a proper transformation
converts the equation into a form solved earlier by Morsh and Feshbach\cite{mf}. Making use
of this solution, we construct the cooperon propagator and 
calculate the dephasing time of a disordered 2DEG in this
modulated field. We find in the weak inhomogeneity limit, $\delta>>L$, where $L$
is the linear size of the sample,
$\tau_\phi^{-1}$ is proportional to $H$, as in the uniform field case. In the
strong inhomogeneity limit, it is shown that 

\begin{equation}
\frac {1} {\tau_\phi}=D\left( \frac e {\hbar c} \right) ^2 H^2\delta^2.
\label{dep2}
\end{equation} 
In the intermediate regime, a
crossover between these two limits is expected to occur at
$H_c=\frac{\hbar c}{4e } \delta^{-2}$.  The dephasing rate dependence on the magnetic field amplitude
is shown to be related to
the nature of corresponding single particle motion. Bound states lead to linear
dependence, while nearly free motion results in quadratic. A semiclassical Monte Carlo
algorithm is developed to study the dephasing time, which is of
qualitative nature but efficient in uncovering the dependence of $\tau_\phi$ on
$H$ for arbitrarily complicated magnetic field modulation. Computer simulations
support analytical results. The considerations are then generalized to situation where
the modulated magnetic field is periodic in one direction with zero mean, and it is
argued that this crossover between linear and quadratic dependence can be expected for a
large class of modulated magnetic fields.

Before going to quantitative calculations, let's see how qualitative
considerations of Khmelnitskii\cite{khm} can predict 
Eq.(\ref{dep2}). Let the magnetic field be nonzero only in a stripe of width
$\delta$. Consider a large loop of area $D\tau$, we are interested in the
flux piercing it. When $\delta$ is small, the effective area where the field is
nonzero is $\delta \sqrt{D\tau}$. Requiring the phase change due to this loop to be
of order 1, $\frac{e}{\hbar c}H \delta \sqrt{D\tau_\phi} \sim 1$, one immediately obtains
Eq.(\ref{dep2}). In subsequent quantitative calculations, we will see that 
understanding of the origin of this dependence allows us to generalize the result
to more complicated situations.

This paper is organized as follows.  In section II, the solution of the
Schr$\ddot{\rm o}$dinger equation by Hud$\acute{\rm a}$k\cite{hud}is briefly
outlined. The construction of cooperon propagator and 
calculation of dephasing time is presented in Section III. Section IV
contains a description of the numerical algorithm and the simulation results. The
generalization to situation with modulated magnetic field periodic in one direction is
presented in Sec.V. The conclusion is given in Section VI.

\section{Solution of the Schr$\ddot{\rm o}$dinger equation 
in ${\mathbf{H}}=(0,0,H/\cosh^2((x-x_0)/\delta))$}

For completeness, the solution of the Schro$\ddot{\rm o}$dinger equation in the 
modulated magnetic field ${\mathbf{H}}=(0,0,H/\cosh^2((x-x_0)/\delta))$ is briefly
outlined in this section. Interested readers are referred to the original article
of Hud$\acute{\rm a}$k\cite{hud} and the book of Morsh and Feshbach\cite{mf}. 

Under the Landau gauge

\begin{equation}
{\mathbf{A}}=(0,H\delta\tanh((x-x_0)/\delta),0),
\end{equation}
the Schr$\ddot{\rm o}$dinger equation can be written as

\begin{eqnarray}
\frac 1{2m} \left[ -\hbar^2 \frac {d^2}{dx^2}+ \left (p_y- \frac {eH\delta}{c}
\tanh ( \frac {x-x_0}{\delta} )\right )^2
\right ] \chi(x) \nonumber \\
=E\chi(x),
\end{eqnarray}
where we have separated variables

\begin{equation}
\psi(x,y)=e^{\frac i \hbar p_y y} \chi(x).
\end{equation}

This Schr$\ddot{\rm o}$dinger equation describes the motion of a particle in the
potential $V(x)$,

\begin{equation}
V(x)=\frac{\hbar^2}{2m\delta^2}\left( 2 \pi \frac {\Phi}{\Phi_0}\right)^2
\left(\frac{p_y \delta/\hbar}{2\pi\Phi/\Phi_0}-\tanh\left(\frac{x-x_0}{\delta} 
\right)  \right)^2
\label{pot2}
\end{equation}
where $\Phi_0=hc/e$ is the flux quantum, $\Phi=H\delta^2$ is a measure of the
magnetic flux by the external field. Introducing

\begin{eqnarray}
F&=&2\pi \frac {\Phi}{\Phi_0}, \nonumber \\
P&=&-p_y\delta/\hbar,\nonumber \\
z&=&(x-x_0)/\delta,
\end{eqnarray}
the potential may be written as

\begin{equation}
V(z)=\frac{\hbar^2 F^2}{2m\delta^2}\left(\frac P F +\tanh z \right )^2.
\end{equation}
For $p_y \ne 0$, this potential is an asymmetric well, with different limiting value
of $V(+\infty)$ and $V(-\infty)$.
Hud$\acute{\rm a}$k\cite{hud} observed that the following transformation 

\begin{eqnarray}
\nu \cosh^2 2\mu &=& P^2+F^2, \nonumber\\
\nu \sinh^2 2\mu &=& 2FP,
\end{eqnarray}
can convert the Schr$\ddot{\rm o}$dinger equation into a form that was solved
by Morsh and Feshbach\cite{mf} earlier,

\begin{eqnarray}
\frac{d^2 \chi(z)}{dz^2}+ \left (\varepsilon-\nu\cosh 2\mu-\nu \sinh 2\mu\tanh z
\right. \nonumber \\
\left.+\nu\frac{ \cosh^2 \mu}{\cosh^2 z} \right )\chi(z)=0,
\end{eqnarray}
where $\varepsilon=2m\delta^2E/\hbar^2$.

One may distinguish situations between $|P| \le F$ and $|P| > F$. In the  former
case, there is
a discrete part in the energy spectrum as well as a continuous one, while the latter
leads to only continuous part. The solution\cite{hud} is as follows. If

\begin{equation}
|P| < F(F-|P|)^2,
\end{equation}
then the energy spectrum for the motion in the $x$-direction contains a discrete
part given by

\begin{eqnarray}
E_n(p_y)=\frac{p_y^2}{2m}\left [ 1-\frac {F^2} {\left (\sqrt{F^2+\frac 1 4
}-(n+\frac 
1 2)\right)^2}\right ] \nonumber \\
+\frac{\hbar^2}{2m\delta^2}\left[ F^2- \left(\sqrt{F^2+\frac 1 4}-(n+\frac 12) 
\right)^2\right],
\end{eqnarray}
with $n=0,1,...,[{\rm n_{max}}]$, and

\begin{equation}
n_{max}=\sqrt{F^2+ \frac14}-\frac 12-\sqrt{|P|F}.
\end{equation}
The corresponding eigenfunction is

\begin{eqnarray}
\chi_n(z)=N_n \exp(-a_nz)(e^{-z}+e^z)^{-b_n}\times F(-n, \nonumber \\
2\sqrt{F^2+1/4}-n,a_n+b_n+1,e^{-z}/(e^{-z}+e^z)),
\end{eqnarray}
where

$$a_n=\frac{|P|F}{\sqrt{F^2+ \frac14}-(n+\frac 12)},$$
$$b_n=\sqrt{F^2+ \frac14}-(n+\frac 12).$$
Here $F(\alpha,\beta,\gamma,\delta)$ denotes the hypergeometric function, and $N_n$
is a normalization constant.

Thus there is discrete part if

\begin{equation}
|P| < P_d(F)=F+\frac1{2F}-\sqrt{1+\frac{1}{4F^2}}.
\end{equation} 

The continuous spectrum may be divided into two parts. For 

\begin{equation}
(F-|P|)^2 < \varepsilon < (F+|P|)^2,
\label{ec1}
\end{equation}
the energy is given by

\begin{equation}
E(k, p_y)=\frac{1}{2m}\left(\left(\frac{\hbar k}{\delta} \right)^2+(|p_y|-p(F))^2 
\right).
\label{ener}
\end{equation}
where $k$ is the momentum in $x$ direction, $p(F)=F\hbar/\delta$. The corresponding
wavefunction is

\begin{eqnarray}
\chi^A_{k,p_y}(z)=N_{k,p_y} \exp(-az)(e^{-z}+e^z)^{-b}\times \nonumber \\
F(b-\gamma+\frac 12,b+\gamma+\frac 12,K_++1,e^{-z}/(e^{-z}+e^z)),
\label{wf1}
\end{eqnarray}
with

$$a=(k_++ik)/2,\qquad b=(k_+-ik)/2,$$
$$\gamma=\sqrt{F^2+\frac 14},\qquad k_+=\sqrt{4F|P|-k^2}.$$
This wavefunction vanishes exponentially for $x \rightarrow \infty$ when $p_y
<0$, and bounded from above when $x \rightarrow -\infty$. For $p_y >0$, the
opposite is true. The inequality (\ref{ec1}) is equivalent to

\begin{equation}
0< k^2 < 4|P|F.
\label{ce2}
\end{equation}

If the parameters are such that the inequality (\ref{ce2}) is replaced by

\begin{equation}
k^2 \ge 4|P|F,
\end{equation} 
then there is another type of continuous energy spectrum. The wavefunctions are
nonvanishing but bounded as $|x| \rightarrow \infty$. They describe overbarrier
motion. The energy is still given by Eq.(\ref{ener}), but the corresponding
wavefunctions are

\begin{eqnarray}
\chi^B_{k,p_y}(z)=N_{k,p_y} \exp(iz(k_+-k)/2)(e^{-z}+e^z)^{i\frac{k+k_+}{2}}
\nonumber \\
\times F((-i(k+k_+)+1-2\gamma)/2,(-i(k+k_+) \nonumber \\
+1+2\gamma)/2,
1-ik_+,e^{-z}/(e^{-z}+e^z)),
\label{wf2}
\end{eqnarray}
where

$$ \gamma=\sqrt{F^2+\frac 14},\qquad k_+=\sqrt{k^2-4F|P|}.$$

The above results are valid for $|P| \le F$. When $|P| > F$, an analytical
continuation can be performed\cite{hud}. And it was found that there is no discrete
part in the spectrum, the wavefunctions and energy are of the same form as that for
$|P| \le F$, but now valid for this region of parameters.

\section{Dephasing time}

In the weak inhomogeneity limit $\delta \gg L$ where $L$ is the linear size
of the sample, $F\gg 1$ for not very weak field. Then the potential well
(given by Eq.(\ref{pot2})) is deep enough
to host many discrete levels, which are reminiscence of the Landau levels. Recall that 
low energy states dominate in the cooperon propagator, the continuous part(with energies higher than 
the barrier height) gives negligible contribution, thus only the
discrete part of the spectrum is important in the cooperon. It can be shown
that in this limit, inhomogeneity brings a correction of order $O(\frac 1 F)$ to
the usual uniform field weak-localization magnetoresistance. And the dephasing time
has the same form with the situation of a uniform field. The opposite limit, $F \ll
1$, which we will focus on, is more interesting.  

Assuming the field described by Eq.(\ref{field}) can be realized in experiment, let's show that
for realistic parameters it is possible to have the discrete part of the spectrum absent in the
strong inhomogeneity limit. 
For $\delta \sim 100 {\rm nm}$, 
$F \sim \frac {H\delta^2} {\Phi_0} \sim 10H$ if $H$ is in Tesla,
thus up to $H\sim 100 {\rm Gauss}$, one can take $F$ as a small quantity. Then we
check if the discrete part of the spectrum exists in this limit. The criterion for
its existence is $P=\frac{p_y\delta}{\hbar}< F^3 < 10^{-3}$.
Since $p_y>\frac{\hbar}{L_\phi}$, one has
$P=\frac{p_y\delta}{\hbar}> \frac {\delta}{L_\phi}$.
For a system with $L_\phi < 10^4 {\rm nm}$, this gives
$P=\frac{p_y\delta}{\hbar} > \frac{10^2 {\rm nm}} { 10^4 {\rm nm}} \sim 10^{-2}$.
Thus the inequality for the existence of discrete levels does not hold in this
situation. One therefore concludes there is no discrete part in the spectrum. In
this limit, 

\begin{equation}
E=\frac{\hbar^2}{2m\delta^2}[p_x^2+(|p_y|-F)^2],
\end{equation}
and the cooperon propagator is 

\begin{eqnarray}
&&C(r,r';\omega)\nonumber \\
&&=\left(\frac {\hbar}{\delta}\right)^2\int dp_xdp_y
\frac{\psi_{p_x,p_y}^*(x,y)
\psi_{p_x,p_y}(x',y')}{-iw\tau+\frac{D\tau}{\delta^2}\left[p_x^2+(|p_y|-F)^2
\right]}, 
\label{coop}
\end{eqnarray}
where the integral should be done under the constraint

\begin{equation}
\frac{D\tau}{\delta^2}\left[p_x^2+(|p_y|-F)^2\right] \ll 1,
\end{equation}
or 

\begin{equation}
p_x^2+(|p_y|-F)^2 \ll q^2, \qquad q^2=\frac{\delta^2}{D\tau},
\end{equation}
as it is a condition for the perturbation theory.

Before proceeding, we need to discuss the boundary condition. 
If we use the wavefunctions Eq.(\ref{wf1}) and (\ref{wf2}) in their 
present form, then the cooperon propagator constructed naively in Eq.(\ref{coop}) 
does not satisfy the zero amplitude boundary condition

\begin{equation}
C(\pm L, y)=C(x, \pm L)=0.
\end{equation}
The boundary condition in $y$ direction can be
taken care of in the same way as in the uniform field case, where the plane wave
wavefunctions are replaced by their linear combinations, and  $p_y$ takes only some
discrete allowed values. The boundary condition in
the $x$ direction can be dealt with in the same manner. We note that from
Eq.(\ref{ener}), the eigenstates with $k$ and $-k$ are degenerate, thus a linear
combination of them is also an eigenstate. The new eigenstate can be constructed as

\begin{equation}
\psi^{A,B}_{p_x,p_y}(x)=\alpha \chi^{A,B}_{p_x,p_y}+ \beta \chi^{A,B}_{-p_x,p_y},
\label{cons}
\end{equation}  
which are required to satisfy the boundary condition

\begin{equation}
\psi^{A,B}_{p_x,p_y}(L)=\psi^{A,B}_{p_x,p_y}(-L)=0.
\end{equation}
As in the case of uniform field, only some discrete $p_x$ will be allowed. Thus the
continuous part of the spectrum becomes again discrete in this boundary condition.
The integration over momentum will be replaced by summation over these discrete
allowed values. We note that the dispersion relation is not altered in the construction
Eq.(\ref{cons}).

Introducing the dimensionless momenta $q_x=p_x/F$, $q_y=p_y/F$, $q'=q/F$, $P'=P/F$,
and going to time domain, we find

\begin{eqnarray}
&&C(r,r';t,t')\nonumber \\
&&=\frac{\hbar^2 F^2 }{\delta^2 \tau}\sum_{q_x^2+(|q_y|-1)^2 \le q'^2}
\psi_{q_x,q_y}^*(x,y)\psi_{q_x,q_y}(x',y') \times \nonumber \\
&&\times
\exp\left\{-\frac{D F^2}{\delta^2}\left[q_x^2+(|q_y|-1)^2\right](t'-t)\right\}
\label{coop2}
\end{eqnarray}
$$=\frac{\hbar^2 F^2 }{\delta^2 \tau}\left\{ \sum_{q_x^2<4|P'|} 
\psi^{A*}_{q_x,q_y}(x,y)\psi^A_{q_x,q_y}(x',y') +\right.$$
$$ \left.+\sum_{q_x^2 \ge 4|P'|} 
\psi^{B*}_{q_x,q_y}(x,y)\psi^B_{q_x,q_y}(x',y')\right\}_{q_x^2+(|q_y|-1)^2 \le q'^2}
$$ 
$$\times 
\exp \left\{-\frac{D
F^2}{\delta^2}\left[q_x^2+(|q_y|-1)^2\right](t'-t)\right\}.$$
We see that the magnetic field results in a characteristic time scale beyond which
the cooperoon propagator or the interference effect is no longer important. Therefore 

\begin{equation}
\frac{1}{\tau_{\phi}}=\frac{DF^2}{\delta^2},
\end{equation}
which leads to Eq.(\ref{dep2}).

Thus the dephasing rate depends quadratically on the field amplitude in the strong inhomogeneity limit. 
In this calculation, we see that the dependence of  $\frac{1}{\tau_{\phi}}$ on $H$ 
is a result of the nature of corresponding single particle
motion. From Eq.(\ref{pot2}), when $F$ is large, the potential well is deep
enough to accommodate many bound states with discrete levels(which are
reminiscence of Landau levels).  Dominant contribution to the cooperon
is from these low lying levels which results in linear dependence. 
In contrast, if $F$ is small, the potential
well can be so shallow that discrete level does not appear, and the particle executes
overbarrier motion which is nearly free with continuous spectrum. When this part
of spectrum is dominant in the cooperon, the dephasing rate is a quadratic function
of the field amplitude. The quadratic dependence implies that the electrons are more slowly
dephased compared with in the uniform field case, where the dependence is linear. 
Physically this is because for a given diffusion time, electrons in this regime
can visit a larger area than the constrained motion. Since the magnetic field is 
nonzero only in a limited region, the phase change accumulated during this time is
smaller than that of the uniform case. So the electron has to wander for a longer time 
before it gets dephased. Therefore the constrained motion of single particle leads
to linear dephasing rate dependence on field amplitude, and nearly free overbarrier motion to quadratic.

Eq.(\ref{dep2}) is obtained in the limit of small $F$, which may be realized with
tiny $\delta$ and small $H$. More realistic for experimental observation is situation with
moderate $\delta$. In this case, one expects a crossover from the linear(uniform limit) 
to quadratic(inhomogeneity limit) dependence. This point can be illustrated in the 
following way. Consider the electron which has diffused for a time $\tau$. If $D\tau<\delta^2$, the
inhomogeneity is not noticed by the electron. If during this time the electron
has already been dephased, $\tau > \tau_{\phi}^u$, where $\tau_{\phi}^u$ is the dephasing time in the
uniform field(given by Eq.(\ref{dep1})), then the inhomogeneity is not important
at all. When $\tau$ is beyond this time scale, the inhomogeneity effect enters. 
Thus, for a given $\delta$, there is a crossover field implied by

\begin{equation}
D\tau_{\phi}^u=\delta^2,
\end{equation}
which yields

\begin{equation}
H_c=\frac{\hbar c}{4e\delta^2}.
\end{equation}
This is equivalent to saying that  the inhomogeneity effect enters when
$F\sim1$. 

On the other hand, if we can adjust the width $\delta$, then for a given $H$,
there is a crossover length scale $\delta_c$,

\begin{equation}
\delta_c=\sqrt{\frac{\hbar c}{4eH}},
\end{equation}
which is the magnetic length for a uniform field.

In passing on, let's mention that from Eq.(\ref{coop2}), one can
see that the magnetoconductance is proportional to $H^2\delta^2$ in this small $F$ 
limit. In the large $F$ limit, a logarithmic dependence is expected as in the
uniform field case. These can also be established by qualitative argument 
of Khmelnitskii\cite{khm}.

\section{Monte Carlo simulation}

A Monte Carlo algorithm has been developed to simulate the dephasing process. In 
order to be consistent  with that implied by Eq. (1),  the simulations will be semiclassical in nature. 
Trajectories will be used and the only quantum mechanical effect will be in the phases. In
this approach, a particle performs random walk in a square lattice. The value of the
perpendicular field is assigned to a dual lattice. We trace all the closed loops
that are formed.  Once a loop is formed, all the inner points are picked up(for technical details, see
[17]), and we calculate and record the
phase change $\phi_i$ due to this loop. Then the trajectory of this loop is erased,
and the particle continues the random walk. The phase accumulated this way is

\begin{equation}
\phi=\sum_i \phi_i,
\end{equation}
whose average is zero since the loop has equal possibility to
be clockwise and counter-clockwise. Then

\begin{eqnarray}
\langle \delta\phi^2 \rangle&=&\langle \phi^2 \rangle \nonumber \\
&=& \sum_i \phi_i^2. 
\end{eqnarray} 
The random walk stops as soon as 

\begin{equation}
\langle \delta\phi^2 \rangle  \ge 1,
\label{cri}
\end{equation}
and we specify the time(total number of random walk steps) that has been spent to
reach this as the dephasing time. An ensemble of such walks are performed and
the dephasing time is averaged. In the simulations, the mean free path
$l$ and the flux quantum $\Phi_0$ have been set to 1, so that 

\begin{equation}
\phi_i=\sum_j H_j,
\end{equation}
where $j$ runs over all the inner points in the dual lattice of loop $i$.

Several features make this simulation of qualitative nature. First, the criterion
Eq.(\ref{cri}) is a qualitative one. Secondly, in the simulation, the impurities are
assumed to be on regular lattice sites, with the mean free path as
lattice constant. Thirdly, sometimes we choose the field not necessarily the 
original one, but qualitatively the same. However, essential physics is not lost despite of
its qualitative nature, and this simulation is very efficient in uncovering {\em
dependence} of the dephasing time on the field amplitude, which is particularly appropriate
for situations where the field modulation is complicated so that it is difficult to
make progress with analytical approaches.

\begin{figure}[tbp]
\centerline{\epsfysize=7.25cm \epsffile{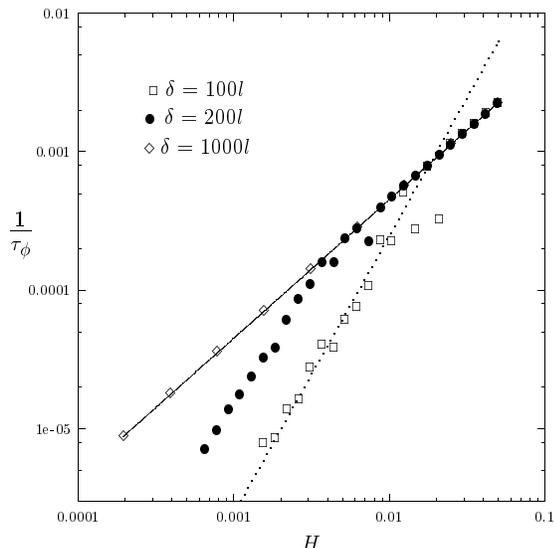}}
\caption{Simulation results for the dephasing time in the modulated magnetic
field Eq.(\ref{fdq1}). The data are obtained by averaging over $10^4$ random
walks. The three curves are for $\delta=1000 l, 200 l, 100 l$,
respectively. The boundary is $L= 5000l$. The straight line is a fit to linear
dependence, dotted line that of quadratic dependence. }
\label{fig1}
\end{figure}

As a check of the algorithm, we have simulated the uniform field case. 
Excellent linearity in the plot of dephasing rate against $H$ is observed, 
in agreement with Eq.(\ref{dep1}). For comparison with the modulated field 
we studied in the previous section, the simulation is performed with the following field

\begin{equation}
H(x)=\left\{ \begin{array}{ll}
H, & |x| \le \delta, \\
0, & {\rm otherwise}, 
\end{array} \right.
\label{fdq1}
\end{equation}
which is qualitatively the same as Eq.(1). The results plotted in
Fig.(\ref{fig1}) show a clear linear
dependence for large $\delta$, and a crossover from linear to quadratic dependence for moderate
$\delta$. The boundary is set at $L= 5000l$ in this simulation. Trajectories which
touch the boundary are excluded. The results are obtained by averaging over
$10^4$ random walks. When the inhomogeneity effect enters,
fluctuations become significant in the result. This is because there are 
extreme trajectories that wander for a long time in the zero field region. However,
by recording not only the average but also the standard deviation, we are able to
get an estimation $2.0 \pm 0.1$ for the exponent in
the inhomogeneous limit, in agreement with Eq.(\ref{dep2}). Thus the numerics
support analytical results.

\section{Generalization to modulated magnetic field periodic in one direction with
zero mean}

The qualitative feature of the results, linear dependence in weak inhomogeneity
limit and quadratic in strong inhomogeneity limit, may be more general. Consider
the situation with magnetic field modulated periodically in one direction with zero
mean. While the weak inhomogeneity limit is easy to understand, we shall focus on
the opposite limit to discuss possible quadratic dependence of $1/\tau_\phi$ on $H$. 

Assume the magnetic-field profile is such that nearest neighbor stripes of 
width $\delta$ have the same magnitude but opposite sign. Consider the phase change
of a closed loop due to this magnetic field. Since
the field is periodic with zero mean, what is important for the magnetic flux is the
ratio of linear size of the loop with the width of one
stripe, $\sqrt{D\tau_\phi}/\delta$. If this ratio is an even integer, then the net
flux is zero. If it is an odd integer, then the flux is equal to that piercing
through a single stripe. In general this ratio may fluctuates, however it is
clear that the average flux is proportional to $\sqrt{D\tau_\phi}H\delta$. Demanding
the phase change due to this flux to be 1, one obtains Eq.(\ref{dep2}), with 
probably an additional numerical coefficient. 

For a general magnetic field periodic in one direction, the corresponding
Schr$\ddot{\rm o}$dinger equation is often difficult to solve analytically. 
For example, for ${\mathbf H}=(0,0,H \sin(x/\delta))$, one can choose ${\mathbf
A}=(0,-H\delta \cos(x/\delta),0)$. Then the resulting Schr$\ddot{\rm o}$dinger
equation is a Whittaker-Hill equation whose solution may be reduced to three-term
recurrence relations\cite{urw}, but an explicit analytical result for the spectrum
is not known. However some insight can be gained by analyzing the structure of energy spectrum. In
general, if the magnetic field is $(0,0,H(x))$, in the Landau gauge ${\mathbf{\rm 
A}}=(0, A(x),0)$.  Separating variables in the wavefunction, the electron motion in
$x$ direction is described by a Schr$\ddot{\rm o}$dinger equation with potential

\begin{equation}
V(x,p_y)=\left [ p_y - \frac{e}{c} A(x) \right ]^2,
\end{equation}
which is also a periodic function. For given $p_y$, this potential is bounded
from above as that described by Eq.(\ref{pot2}). The profile of this potential  is a
series of potential wells joining with each other.
The energy spectra of electrons in a family of such modulated magnetic fields
are calculated by Ibrahim and Peeters\cite{ip}. For electron energy less than the
barrier height, the wavefunctions in neighboring wells overlap and
spread. So the discrete levels in single wells now form  minibands due to the
periodic structure. When $H$ and $\delta$ are large, the barrier can be quite
high, and states in these minibands dominate in the
cooperon propagator. According to the experience in dealing with the previous case,
the dephasing rate is expected to be a linear function of $H$. When $H$ and $\delta$
are small, however, the potential well is not deep enough to support these
minibands, and dominant contribution is expected from nearly free 
overbarrier motion, which leads to a quadratic dependence.

\begin{figure}[tbp]
\centerline{\epsfysize=7.25cm \epsffile{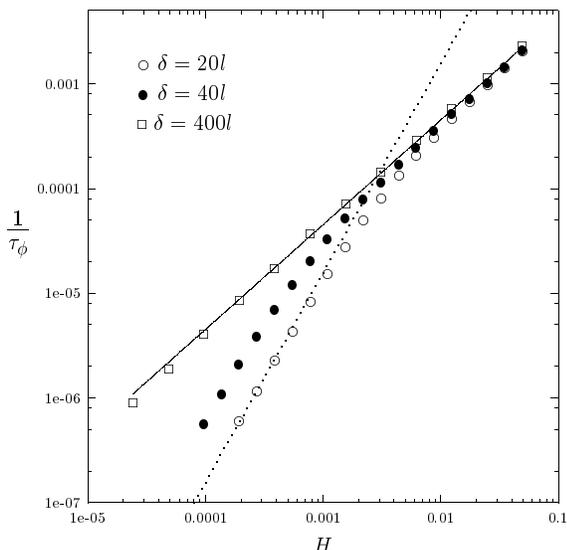}}
\caption{Simulation results for the dephasing time in the modulated magnetic
field periodic in one direction given by Eq.(\ref{fdq2}). The data are obtained by
averaging over an ensemble of $10^3$ random walks. The three curves are for
$\delta=400 l, 40 l, 20 l$, respectively.  The straight line is a fit to linear
dependence, the dotted line that of quadratic dependence.}
\label{fig2}
\end{figure}

To examine these considerations, computer simulations are performed for the
following field

\begin{equation}
H(x)=\left\{ \begin{array}{ll}
H, & 2n\delta \le x < (2n+1)\delta, \\
-H, & (2n-1) \delta \le x < 2n\delta. 
\end{array} \right.
\label{fdq2}
\end{equation}
The results are shown in Fig.\ref{fig2}. The curves show much resemblance to that
in Fig.\ref{fig1}, and a crossover from linear to quadratic dependence is evident. 
There are also some differences which will be noted
here. First, for a given $\delta$, the crossover field $H_c$ is smaller in the field
(\ref{fdq2}) than in (\ref{fdq1}), which suggests when $H$ is small, electron in the
field (\ref{fdq2}) is more quickly dephased than in the field (\ref{fdq1}). This may
look strange at first sight, since the field
(\ref{fdq2}) has zero mean, one may expect the electrons in this field are
more slowly dephased. However, this result  can be understood  by
noticing what enters is the variance of the phase instead of the average, and the
field (\ref{fdq2}) is nonzero everywhere in the plane. Another difference is in the
field (\ref{fdq2}) there is broader crossover region. This is associated with the
fact that, in the periodic situation, states with energy lower than the barrier
height are not really bound states. They share some feature of the plane waves
according to Bloch's theorem. As discussed previously, this feature has the 
tendency of leading to a quadratic dependence. When $\delta$ is moderate, 
the barrier is not very high for many values of $p_y$. Then these states 
can be dominant in the cooperon, resulting in a broad crossover region.

Numerical experiments are also performed for situations with some other
magnetic field modulations, for example, that with circular symmetry, or periodic in
both directions. Similar crossover behaviors have been observed. These simulations
as well as the qualitative understanding of the origin of the dependences, suggest
that the  crossover from a linear to quadratic dependence in the dephasing rate on
field amplitude can be quite
general. For situations where the corresponding single particle potential is bounded
from above, the particle motion can exhibit both bound and free nature, leading
to a crossover from linear to quadratic dependence. 

\section{Conclusion}

We have studied the dephasing time of disordered two-dimensional electron gas in
modulated magnetic field ${\mathbf{H}}=(0,0,H/\cosh^2((x-x_0)/\delta))$. It is
shown that in
the weak inhomogeneity limit, $\delta>>L$, where $L$ is the linear size of the
sample, $\tau_\phi^{-1}$ is proportional to $H$. This happens when the bound 
states with discrete spectrum of the corresponding Schr$\ddot{\rm o}$dinger
equation dominate in the cooperon propagator. While in the strong inhomogeneity
limit, the dependence is quadratic  $\tau_\phi^{-1}=D\left( \frac e {\hbar c}
\right) ^2 H^2\delta^2$. In this case, the nearly free overbarrier motion gives 
dominant contribution to cooperon. In the intermediate regime, a crossover between
these two limiting situations occurs at $H_c=\frac{\hbar c}{4e}
\delta^{-2}$. A semiclassical Monte Carlo algorithm has been developed to study the dephasing
time, which is of qualitative nature but efficient in uncovering the dependence of
$\tau_\phi$ on $H$ for arbitrarily complicated magnetic field modulations. Computer 
simulations support analytical results. These considerations are generalized to the situation
with magnetic field modulated periodically in one direction with zero mean, where a
similar crossover is observed. We believe
this crossover between linear and quadratic dependence can be expected for a large
class of modulated magnetic fields where the corresponding single particle
potential is bounded from above, so that the motion
exhibits both bound and free nature depending on parameters.

\vskip 1cm
\centerline{\bf \large Acknowledgements}
\vskip 0.5cm

I am grateful to Vladimir Kravtsov for helpful discussions. Discussions on the
simulation algorithm with Dr. Z. F. Huang, Dr. F. Ricci-Tersenghi, and a
communication with Prof. K. Kremer and Prof. K. Binder are acknowledged.

\end{multicols}

\end{document}